\documentclass{aastex}
\shorttitle{A solar dynamo model}
\usepackage{emulateapj5}
\newcommand{\p}[2]{\frac{\partial #1}{\partial #2}}

\newcommand{\vb}[1]{{\bf #1}}
\newcommand{\s}[1]{\mbox{\scriptsize{#1}}}

\begin{document}

\title{A convection zone dynamo including the effects of magnetic buoyancy and downward flows}

\author{L. H. Li\altaffilmark{1} and S. Sofia}
\affil{Department of Astronomy, Yale University, P.O. Box 208101, New Haven, CT 06520-8101}
\altaffiltext{1}{also at Purple Mountain Observatory,  Chinese Academy of Sciences, and
National Astronomical Observatories, Chinese Academy of Sciences, 2 Beijing Road West, Nanjing,
Jiangsu 210008, China}
\email{li@astro.yale.edu}
\email{sofia@astro.yale.edu}

\begin{abstract}
A magnetic flux tube may be considered both as a separate body and as a confined field. As a field, it is affected 
both by the cyclonic convection ($\alpha$-effect) and differential rotation ($\Omega$-effect). As a body, the tube experiences not only a buoyant force, but also a dynamic pressure due to downflows above the tube. When these two dynamic effects are incorporated into the $\alpha\Omega$ dynamo equations, we obtain a dynamo operating in the convection zone. We analyze and solve the extended dynamo equations in the linear approximation by using observed solar internal rotation and assuming a downflow suggested by numerical simulations of the solar convection zone. The results produce:
(i) the 22-year cycle period; (ii) the extended butterfly diagram; (iii) the confinement of strong activity to low heliographic latitudes $|\Phi|\le 35^\circ$; (iv) at low latitudes the radial field is in an approximately $\pi$ phase lag 
 compared to the toroidal field at the same latitude; (v) the poleward branch is in a $\pi/2$ phase lag with respect to the equatorward branch; (vi) most of the magnetic flux is present in a strongly intermittent form, concentraed 
into strong flux tubes; (vii) the magnetic field peaks at a depth of $r=0.96 \,R_{\sun}$;
(viii) total solar irradiance varies in phase with the solar cycle activity, having an amplitude of $0.1\%$;
(ix) solar effective temperature varies in phase with the solar cycle activity, having an amplitude of 1.5 $^\circ C$;
and (x) solar radius also varies in phase with the solar cycle activity, having an amplitude of 20 mas. All these results are in agreement with the corresponding observations.
\end{abstract}
\keywords{Sun: interior --- Sun: magnetic fields}

\maketitle

\section{Introduction}\label{sect:s0}

Since Parker set down the foundations for solar $\alpha\Omega$ 
dynamo theory in his classic 1955 paper, many solar cycle models 
have been proposed, for example, the convection zone (CZ) dynamo 
models \cite{P55}, the overshoot layer (OL) dynamo models 
(e.g., R\"{u}diger \&\ Brandenburg 1995), the interface 
(IF) dynamo models \cite{P93,CM97,MT99}, and the Babcock-Leigton (BL) 
dynamo models \cite{CSD95,D95,D96,D97,NC01}. In CZ dynamos, 
differential rotation (the $\Omega$-effect) in the solar interior 
shears weak poloidal magnetic fields into intense toroidal fields. 
Meanwhile twisting motions associated with buoyancy, turbulent 
convection, and the Coriolis force transform toroidal fields back 
into poloidal fields (the $\alpha$-effect). Since most of solar 
magnetic flux is present in flux tubes and they are subject to 
magnetic buoyancy instabilities, if there is no force to balance 
the buoyancy, they cannot be stored in the convection zone for a 
timescale comparable to the cycle period \cite{P75}.

In order to circumvent the difficulty of flux storage in the main bulk 
of the convection zone, OL dynamos were proposed. Observations (Fig.~\ref{fig:rot}) in fact show 
that most of the shear (the $\Omega$-effect) is concentrated in a thin layer near the 
bottom of the convection zone, known as the tachocline. This was seen as evidence 
that the tachocline is of key importance for the working of the solar dynamo. Such 
dynamos, use the same concept as CZ dynamos. In such dynamos, 
the butterfly diagram comes out right, but the radial field is found to be nearly in phase 
with the toroidal field, instead of being in antiphase as observed, and the cycle period 
tends to be too short for thin layer models.

Unlike both CZ and OL dynamos, in which 
the $\alpha$- and $\Omega$-effects are assumed to be cospatial, in IF dynamos 
the $\alpha$-effect 
operates on the high-diffusivity side of an interface, while the $\Omega$-effect is 
limited to the low-diffusivity side. Their attractive feature is that the 
toroidal field generated on the low diffusivity side can 
be made arbitrarily strong by reducing the value of magnetic diffusivity there. 
The base of the convection zone, where the convection 
zone and the radiative zone are separated naturally, provides a physical interface.

Unlike CZ, OL, and IF dynamos, which interpret the butterfly diagram as a 
dynamo wave, BL dynamos essentially work like a conveyor belt: 
the poleward meridional circulation near the surface transports the poloidal fields 
towards the poles at high latitudes, giving rise to the poleward branch of the butterfly 
diagram. At the poles, the fields are advected down to the bottom of the convection 
zone where the shear converts them into toroidal fields that get amplified while advected 
towards the equator. Once these are strong enough, they are supposed to form buoyantly 
emerging loops at low latitudes that give rise to active regions. The Coriolis force lends 
an inclination to the loop planes, introducing an $\alpha$-effect to regenerate the poloidal 
field. In such models, the low-latitude confinement of strong activity comes out naturally 
since the toroidal field is amplified by shear as it is advected equatorwards, and the 
extended butterfly diagram is reproduced approximately in the sense that the latitude 
where the two branches part tends to be too high. A serious problem is that the approach 
only works with an unrealistically low value for the turbulent diffusivity in the convection zone 
in order to keep the two parts of the conveyor belt separated.

A successful solar cycle model should interpret the following observational facts:
\begin{enumerate}
 \item The 22-year cycle period;
 \item The characteristic migration pattern (extended butterfly diagram, e.g., Makarov 
  \&\ Sivaraman 1989);
 \item The confinement of strong activity to low heliographic latitudes $|\Phi|\le 35^\circ$;
 \item The phase dilemmas \cite{S76}:
  \begin{enumerate}
   \item at low latitudes the radial field is in an approximately $\pi$ phase lag 
 compared to the toroidal field at the same latitude;
   \item the poleward branch is in a $\pi/2$ phase lag with respect to the equatorward branch; 
  \end{enumerate}
 \item Most of the magnetic flux is present in a strongly intermittent form, concentraed 
into strong flux tubes \cite{W64,S66,S73,HS72}.
\end{enumerate}
Helioseismic observations also impose rigorous constraints on the model:
\begin{enumerate}
 \setcounter{enumi}{5}
 \item Solar internal differential rotation rate from inversions of helioseismic observations 
(e.g., Antia et al. 1998) should be used in the model;
 \item Helioseismic inversions made by Antia, Chitre, and Thompson (2000) indicate 
that there is a magnetic field of approximately 20 kG strength peaked at a depth of 
$r=0.96 \,R_{\sun}$ in 1996, rather than at the base of the convection zone.
\end{enumerate}
Furthermore, the model should reproduce the observed cyclic variations of total solar irradiance, solar effective temperature, and solar radius:
\begin{enumerate}
 \setcounter{enumi}{7}
 \item Total solar irradiance varies in phase with the solar cycle activity, having an amplitude of $0.1\%$ \cite{FL98};
 \item Solar effective temperature varies in phase with the solar cycle activity, having an amplitude of $1.5^\circ C$ \cite{GL97};
 \item Solar radius also varies in phase with the solar cycle activity, having an amplitude of 20 mas \cite{EKBS00}.
\end{enumerate}
Using these constraints, we can see that none of the above solar dynamos are 
satisfactory. In particular, criterion 7 rules out the last three models since 
they predict that solar magnetic fields peak at the base of the convection zone, 
leaving CZ dynamos as the unique candidate.

Recently, Li and Sofia (2001) have shown that a magnetic field of $20-50$ kG 
located at $r=0.96 R_\sun$ can reproduce the observed cyclic variations of total 
solar irradiance, solar effective temperature, and solar radius. If we use the 
helioseimic internal differential rotation in the CZ dynamo, the CZ dynamo 
will automatically satisfy criteria~$6-10$. In order to meet criteron 5,
the magnetic fields should be considered in the form of strong flux tubes. As a 
body, a flux tube in the convection zone experiences not only a magnetic 
buoyancy since gas in it is less dense than the surroundings \cite{P75}, but also 
a dynamic pressure if flow is present. Stratification provides a 
preferred direction; turbulence in the convection zone is likely to be anisotropic.
In fact, numerical simulations of the solar outer convection zone \citep{CS89,KC98,N99}
indicate a major presence of downward-moving plumes. This suggests that the dynamic 
pressure of the downward-moving flow may push down the magnetic flux tubes 
to prevent them from rising, until they reach certain depth in the convection zone.

In this paper, we extend the classical CZ dynamo model by taking into account the 
magnetic buoyancy and downward flow effects. We show in \S\ref{sect:s1}
that these two effects can be incorporated into the mean field dynamo equations by modifying
the turbulent diffusivity $\beta$. In \S\ref{sect:s2} we linearly analyze the dynamo 
equations to obtain the observable quantities of the model. In \S\ref{sect:s3} we use the 
theoretical findings to interpret the observational facts listed above. Finally, 
in \S\ref{sect:s4} we summarize the results.

\section{Basic equations}\label{sect:s1}

We consider a magnetic flux tube with radius $a$ and length $L$. Its magnetic field 
$\vb{B}$ is assumed to be uniform. As usual, we decompose it into a toroidal and a 
poloidal component,
\begin{equation}
  \vb{B} = B \vb{e}_{\phi} + \nabla\times (A\vb{e}_{\phi}), \label{eq:b}
\end{equation}
where $\vb{e}_\phi$ is the azimuthal unit vector, $B$ is the toroidal field component, and $A$ is the (toroidal) vector potential of the poloidal field. The local hydrostatic equilibrium of the flux tube in the solar interior requires
\begin{equation}
P = P_i+B^2/8\pi,
\end{equation}
where $P$ and $P_i$ are the total and internal gas pressure, respectively. Assuming 
that this equilibrium in the flux tube is reached solely by adjusting its density, 
then the internal density $\rho_i$ is related to the normal density $\rho$ by
\begin{equation}
  \rho_i = \rho P_i/P.
\end{equation}
As a result, the density reduction of the magnetic flux tube is
\begin{equation}
 \rho-\rho_i=\rho B^2/8\pi P.
\end{equation}
The reduced density inside the flux tube produces a buoyancy,
\begin{equation}
  \vb{F} = \pi a^2L g(\rho-\rho_i) \vb{e}_r,
\end{equation}
where $g$ is the gravitational acceleration. We assume that $a$ is smaller than 
the pressure scale height so that we can consider that the gas density of the 
flux tube is uniform. In this case, the total mass contained in the flux tube equals
\begin{equation}
  M_i = \pi a^2L\rho(1-B^2/8\pi P).
\end{equation}
Therefore, the buoyant acceleration $g_b$ can be expressed by the gravitational 
acceleration $g$, total pressure $P$ and the magnetic field strength $B$ as follows:
\begin{equation}
  g_b = gB^2/(8\pi P-B^2). \label{eq:gb}
\end{equation}

If there is no downward flow to balance this pressure, the only force that goes 
agaist the buoyant force is the aerodynamic drag,
\begin{equation}
  \vb{F}_d = -\case{1}{2}C_d\rho u^2 aL \vb{e}_r,
\end{equation}
where $u$ is the velocity of rise and $C_d (\sim 1)$ is the drag coefficient. 
Taking into account both the buoyant force and the drag force, as done by 
Parker (1975), the terminal velocity of rise occurs for $F=F_D$, yielding 
the rate of rise
\begin{equation}
   u = \left(\frac{2\pi a g}{C_d}\frac{\rho-\rho_i}{\rho}
  \right)^{1/2}=V_A\left(\frac{\pi a}{C_D H_p}\right)^{1/2},
\end{equation}
where $H_p=P/\rho g$ is the pressure scale height, $V_A=B/(4\pi\rho)^{1/2}$ 
is the Alfv\'{e}n speed. The rise time to the surface is smaller than the 
required field amplification time of about 10 years by the classical dynamo 
theory, as estimated by Parker (1975).

However, numerical simulations of the Sun's outer convection zone made, 
e.g. by Nordlund (1999), indicate a major presence of downward-moving 
plumes of high velocity. This downward flow has an inward dynamic pressure 
$\rho v_z^2$. If this pressure can balance the buoyant force of the flux 
tube, then the flux tube can be stored in the convection zone so that it 
can be amplified by the $\alpha\Omega$ dynamo. Antia et al (2000) employ 
the observed splittings of solar oscillation frequencies to separate the 
effects of interior solar rotation, and to estimate the contribution from 
a large-scale magnetic field. After subtracting out the estimated 
contribution from rotation, there is some residual signal in the even 
splitting coefficients. This may be explained by a magnetic field of 
approximately 20 kG strength located at $r=0.96 R_{\sun}$ in 1996. 
Since the density here is of order $4\times10^{-3}$, and the downward 
velocity for the plumes is of order $5\times10^4$ cm s$^{-1}$ \citep{N99},
the estimated dynamical pressure of the plumes, $\rho v^2$, is equal to or 
larger then $10^7$ dyne cm$^{-2}$. The size of $\rho v^2$ is comparable 
with the magnetic pressure, $B^2/8\pi$, for a field strength of $20-30$ 
kG. Therefore, as a body, a magnetic flux tube experiences not only a 
buoyant force, but also a dynamic pressure of downflows. The total 
pushdown force of downward plumes equals to the dynamic pressure 
($\rho v_z^2$) times the total cross section of the plumes ($S$). 
The dynamic acceleration $g_n$ can be expressed as follows:
\begin{equation}
  g_n = 2s v_z^2 \xi/\pi a, \label{eq:gn}
\end{equation}
where $s=1$ when $v_z>0$, $s=-1$ when $v_z<0$, $\xi=S/2aL\le 1$ is the 
fractional area of the downflows. The upward-moving plumes underneath 
the flux tube ($s=1$) accelerate the magnetic buoyant diffusion of the 
tube, while the downward-moving plumes above the flux tube ($s=-1$) go 
against the magnetic diffusion. These downflows act as an anti-diffusion 
process.

The magnetic field in a flux tube can be considered to be a ``mean field'' 
since the mean-field concept is relative in Mean Field Magnetohydrodynamics (e.g., 
Krause and  R\"{a}dler 1980). As a matter of fact, it depends on the 
spatio-temporal range over which the average is made. For the mean field 
in a flux tube, the spatio-temporal range is the volume and lifetime of 
a typical magnetic flux tube. With this in mind, the field in a flux tube
obeys the classic dynamo equation \cite{P55},
\begin{equation}
  \p{\vb{B}}{t} = \nabla\times(\vb{U}\times\vb{B} + \vec{{\cal E}}) -\nabla\times( \eta\nabla\times\vb{B}). \label{eq:dynamo}
\end{equation}
Using this equation, we cannot study the formation of flux tubes from the 
large-scale magnetic field, but we can study if the flux tube fields can be
amplified.

We assume a pure rotation motion $\vb{U}=\vb{\Omega}\times\vb{r}$, where $\vb{\Omega}$ is the angular velocity.
This will cause the $\Omega$-effect. $\vec{{\cal E}}$ contains the $\alpha$-effect, $\alpha \vb{B}$, and the magnetic diffusion processes, which are caused by the turbulent pressure, $\beta_t\nabla\times\vb{B}$, by the magnetic buoyant force, $\beta_b\nabla\times\vb{B}$, and by the dynamic pressure of downward-moving plumes, $\beta_n\nabla\times\vb{B}$. Therefore,
\begin{equation}
 \vec{\cal E} = \alpha \vb{B} - (\beta_t+\beta_b+\beta_n) \nabla\times \vb{B}.
\end{equation}
The magnetic diffusion caused by the microscopic gas collision motion, has been included in Eq.~(\ref{eq:dynamo}) in terms of the classic magnetic diffusivity $\eta\nabla\times\vb{B}$. For convection, $\alpha$ can be approximately expressed as \citep{KR80}
\begin{equation}
  \alpha = \case{1}{3} h \tau_{\s{cor}},
\end{equation}
where $h=\overline{\vb{v}\cdot(\nabla\times\vb{v})}$ is the mean helicity over the correlation time $\tau_{\s{cor}}$.

\vspace{3mm}
\centerline{\epsfysize=8.cm \epsfbox{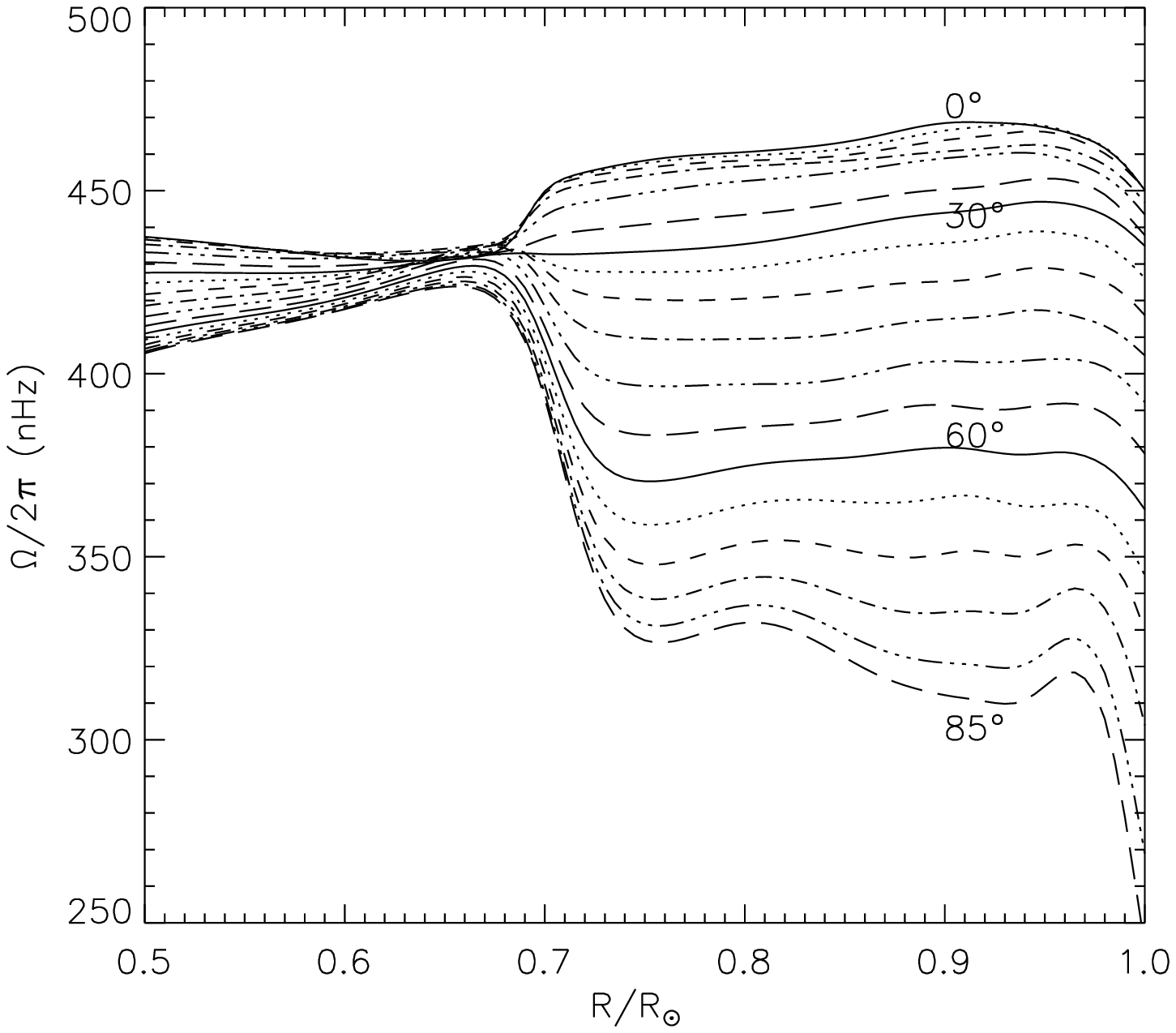}}
\figcaption[f1.eps]{
Solar rotation rate as a function of radius and latitude \citep{ABC98}.
\label{fig:rot}}
\vspace{3mm}

In order to work out the various magnetic diffusitivites mentioned above, we observe that the diffusivity of a force is proportional to its impulse per unit mass. The impulse is defined as the force times its action time. The impulse per unit mass has the dimension of acceleration times time. The action time can be taken to be the propagation time of a magnetic perturbation across the diameter of the flux tube. Since the characteristic propagation velocity of magnetic perturbations is the Alfv\'{e}n velocity $V_A$, the action time is estimated to be $t_{\s{act}}=2a/V_A$. However, diffusivity has the dimension of acceleration times time times length. The charateristic length for a flux tube is its length $L$. We use this length $L$ to meet the need of dimensionality for the buoyant and the dynamic diffusivity. Consequently, we have
\begin{eqnarray}
 \beta_b &=& g_b (2a/V_A)L=(4\pi \rho)^{1/2} 2agL B/(8\pi P-B^2) , \label{eq:betab} \\
 \beta_n &=& g_n (2a/V_A)L= (4\pi\rho)^{1/2}4sv_z^2\xi L/\pi B. \label{eq:betad}
\end{eqnarray}
The magnetic turbulent diffusivity $\beta_t=\case{1}{3} v'^2 \tau_{\s{cor}}$ is known \citep{KR80} for an isotropic turbulence, where $v'$ is the turbulent velocity.

In summary, we define
\begin{equation}
  \beta = \eta + \beta_t+\beta_b+\beta_n,
\end{equation}
where $\beta_b$ and $\beta_d$ are given by (\ref{eq:betab}) and (\ref{eq:betad}), respectively. As a result, the hydromagnetic induction equation becomes the dynamic equation of a magnetic flux tube,
\begin{equation}
  \p{\vb{B}}{t} = \nabla\times(\vb{U}\times\vb{B}+\alpha\vb{B}) -\nabla\times( \beta\nabla\times\vb{B}). \label{eq:model}
\end{equation}
Since $\beta$ depends on $\vb{B}$, this equation is nonlinear.

\section{Linear analysis}\label{sect:s2}

The flux tube model equation obtained in the previous section is a nonlinear partial-differential vector equation. Since such equation is difficult to solve analytically, while a simple linear analysis may reveal some important characteristics about its solution, we analyze and solve the equation in the linear approximation in this paper. Unlike numerical simulations, the linear analysis cannot take into account full nonlinearity though it reveals some nonlinear effects such as critical magnetic fields of a flux tube.

\subsection{Linearization of equations}

Substituting Eq.~(\ref{eq:b}) into Eq.~(\ref{eq:model}), and linearizing the result with respect to the perturbation of $B$ and $A$ near $B_0$ and $A_0=0$, denoted by $B'$ and $A'$, we obtain two linear equations that govern $B'$ and $A'$,
\begin{eqnarray}
  \p{A'}{t} &=& \alpha B' + \beta_0 \nabla^2 A', \label{eq:linear1}\\
  \p{B'}{t} &=& \hat{\Omega}_z \p{A'}{x}-\hat{\Omega}_x\p{A'}{z} -\alpha \nabla^2 A' + \beta_0\nabla^2B' \nonumber\\
     && +\hat{\beta}_B B'-\hat{\beta}_z\p{B'}{z} +\hat{\beta}_x\p{B'}{x}  \label{eq:linear2}
\end{eqnarray}
in the local Cartesian frame $\vb{e}_z=\vb{e}_r$, $\vb{e}_y=\vb{e}_\phi$, $\vb{e}_x=\vb{e}_{\theta}$. We have defined,
\begin{eqnarray}
  \hat{\Omega}_x&=&\p{}{x}(r\Omega\sin\theta)=\Omega(\cos\theta+\hat{\Omega}_\theta \sin\theta), \nonumber\\
  \hat{\Omega}_z&=&\p{}{z}(r\Omega\sin\theta)=\Omega\sin\theta(1+\hat{\Omega}_r), \nonumber\\
  \hat{\beta}_B &=& \hat{\beta} \nabla^2B_0-\p{\hat{\beta}}{z}\p{B_0}{z}+\p{\hat{\beta}}{x}\p{B_0}{x}, \\
  \hat{\beta}_x &=& \p{\beta_0}{x}+\hat{\beta}\p{B_0}{x}, \nonumber\\
  \hat{\beta}_z &=& \p{\beta_0}{z}+\hat{\beta}\p{B_0}{z}, \nonumber
\end{eqnarray}
where $\beta_0=\beta(B_0,v_z)$, $\hat{\Omega}_\theta=\partial{\ln\Omega}/\partial{\theta}$, $\hat{\Omega}_r=\partial\ln\Omega/\partial \ln r$, and $\hat{\beta}=(\partial\beta/\partial B)|_{B=B_0}=[\beta_1(B_0)-\beta_2(B_0,v_z)]/B_0$. The second line of Eq.~(\ref{eq:linear2}) stands for the nonuniform effect. When this effect is ignored, 
Eqs.~(\ref{eq:linear1}-\ref{eq:linear2}) are the same as the classic dynamo equations, except that $\beta_0$ depends on the magnetic field and the convection velocity field $v_z$. In the classic case, $\beta_0$ is always positive, implying a genuine diffusion. In contrast,
when the dynamic pressure of the downward moving plumes is larger than the buoyant pressure of the flux tube,
 $\beta_0$ can become negative.

\subsection{Dispersion relation}

In order to investigate the linear instability, we try the wavelike solutions of the form
\begin{eqnarray}
  B' &=& B'_0\exp[i(\omega t-\vb{k}\cdot\vb{x})], \label{eq:solve1}\\
  A' &=& A'_0 \exp[i(\omega t-\vb{k}\cdot\vb{x})], \label{eq:solve2}
\end{eqnarray}
where $\omega$ is complex, while the other variables are real. Substituting Eqs.~(\ref{eq:solve1}-\ref{eq:solve2}) into Eqs.~(\ref{eq:linear1}-\ref{eq:linear2}), we obtain
\begin{eqnarray}
  (\beta_0 k^2+i\omega)A'_0 &=& \alpha B'_0, \label{eq:disp1}\\
  (\beta_0 k^2+i\omega-\hat{\beta}_B+i\hat{\beta}_k)B'_0 &=& -i(k_x\hat{\Omega}_z-k_z\hat{\Omega}_x) A'_0,  \label{eq:disp2}
\end{eqnarray}
where $\hat{\beta}_k=\hat{\beta}_xk_x-\hat{\beta}_zk_z$.
The dispersion equation can be obtained by multiplying Eq.~(\ref{eq:disp1}) and (\ref{eq:disp2}),
\begin{equation}
  (\beta_0 k^2+i\omega)(\beta_0 k^2+i\omega-\hat{\beta}_B+i\hat{\beta}_k) = -i \alpha (k_x\hat{\Omega}_z-k_z\hat{\Omega}_x).
    \label{eq:relation}
\end{equation}
Solving this equation for $\omega$ as a function of wavenumber vector $\vb{k}$, we obtain the dispersion relation,
\begin{equation}
  \omega=\omega(k_x,k_y,k_z;v_z,B_0),
\end{equation}
where $B_0$ and $v_z$ are parameters.

Under the homogeneity approximation, we can neglect $\hat{\beta}_B$ and $\hat{\beta}_k$. In this case, Eq.~(\ref{eq:relation}) reduces to
\begin{equation}
  (\beta_0 k^2+i\omega)^2 = -i \alpha (k_x\hat{\Omega}_z-k_z\hat{\Omega}_x).
\end{equation}
Rewriting $-i$ as $\exp(i3\pi/2)$, when
\begin{equation}
 D=\alpha(k_x\hat{\Omega}_z-k_z\hat{\Omega}_x)/2\ge 0 \label{eq:d}
\end{equation}
is satisfied, we can solve this equation for $\omega$ easily as follows:
\begin{equation}
  \omega^{\pm}=\pm\sqrt{D} + i(\beta k^2\mp\sqrt{D}). \label{eq:omega}
\end{equation}

\subsection{Period}

When $D\ne 0$, the magnetic flux tube generation is periodic. The period is
\begin{equation}
   T=2\pi/\sqrt{D}. \label{eq:period}
\end{equation}
It is well-known that $\alpha$ is positive (negative) for cyclonic convection in the northern 
(southern) hemisphere of the Sun. Therefore, the condition $D>0$ leads to
\begin{equation}
  (1+\hat{\Omega}_r) k_x> (\cot\theta + \hat{\Omega}_\theta) k_z.
\end{equation}
When this condition is satisfied we have two magnetic flux tube waves propagating in the opposite dirrection. These two waves have the same period.

In order to calculate $T$, we first determine $k_x$ and $k_z$. The fundamental wavenumbers are, $k_{x0}=\sqrt{2}\,\pi/r$ in the $x$-axis, and $k_{z0}=2\pi/r$ in the $z$-axis at $r$ since the maximum wavelengths are $\lambda_{x0}=\sqrt{2}\,r$ and $\lambda_{z0}=r$. The butterfly diagram implies that both branches share the whole length, so we assume $k^\pm_x=\pm k_{x0}$ for the equatorward/poleward branch in the northern hemisphere of the Sun. The condition $D>0$ leads to $k^\pm_z=\pm k_{z0}$ in the same hemisphere.

Helioseismic data allow us to probe the rotation rate in the solar interior as a function of radius and latitude \citep{TET96,ABC98,SET98,AB00}. We use the rotation rate inferred from Global Oscillation Network Group (GONG) data by Antia, Basu, and Chitre (1998), as shown in Fig.~\ref{fig:rot}.

Table~\ref{tbl:period} shows $T$ (unit: year) as a function of radius and latitude, calculated by using Eq.~(\ref{eq:period}). This table shows a low latitude branch and a high latitude branch. Since the observed length of the solar cycles is equal to about 22 years, this table shows that the cycle should originate from the surface layer with $r>0.95 R_{\sun}$, which is in agreement with the inferred depth from helioseismic data by Antia, Chitre and Thompson (2000), who found that the solar magnetic field peaks at $r=0.96 R_{\sun}$. The period also depends on latitude. If we use the observed period (22 years) as a criterion, this table indicates that the low- (high-) latitude branch originates from  the latitude 35$^{\circ}$ (40$^{\circ}$). Since $k_x>0$ and $k_z>0$ for the low-latitude branch, the magnetic flux tubes  near the latitude 35$^{\circ}$ and radius $r=0.96\,R_{\sun}$ move upwards and equatorwards. Since $k_x<0$ and $k_z<0$ for the high-latitude branch, the magnetic flux tubes near 40$^{\circ}$ and $0.96 R_{\sun}$ move downwards and polewards. This scenario is in agreement with observations \cite{MS89}.

For different branches we have to use different $\alpha$ values to produce the desired period at the mid latitude. Since $\alpha$ is determined by the Coriolis force and the higher the Coriolis force the lower the latitude, it is reasonable to use a larger $\alpha$ for the low-latitude branch. Nevertheless, a jump of $\alpha$ from $90$ cm s$^{-1}$ at 35$^\circ$ to 5 cm s$^{-1}$ at 40$^{\circ}$ is implied and needs to be explained.

Table~\ref{tbl:period} also shows that solar magnetic activity is a multi-period phenomenon, and the longest period is 22 years, occuring in the latitude range from 35$^{\circ}$ to 40$^{\circ}$. The period decreases equatorwards, polewards, and downwards. The period can be as small as 1-2 year near the base of the convection zone and near the equator.

\subsection{Critical magnetic field}

Eqs.~(\ref{eq:solve1}-\ref{eq:solve2}) show that spontaneous generation of a magnetic flux tube requires $\omega_I=(\beta_0 k^2\mp\sqrt{D})<0$, which implies 
\begin{equation}
  \beta_0<\pm\sqrt{D}/k^2. \label{eq:unstable}
\end{equation}
For simplicity, we further neglect the classic collision and isotropic turbulent diffusivity. As a result, the unstable condition (\ref{eq:unstable}) becomes 
\begin{equation}
 \beta_b+\beta_n\mp\sqrt{D}/k^2 <0. \label{eq:criterion}
\end{equation}

We assume $B_0\ll B_g$, where $B_g=(8\pi P)^{1/2}$ is the magnetic field strength when the magnetic pressure equals to the total pressure $P$. In this case inequality~(\ref{eq:criterion}) is quadratic since $\beta_b\approx(4\pi \rho)^{1/2} 2agL B/B_g^2$. Solving this quadratic inequality for $B_0$, we obtain
\begin{equation}
  B^+_-<B_0<B^+_+
\end{equation}
for the $\omega^+$ branch, where
\begin{eqnarray}
B^+_\pm &=& B_g\left[\left(\frac{DP}{8\rho a^2L^2g^2k^4} \right)^{1/2} \right. \nonumber \\
 &&  \pm \left.\left(\frac{DP}{8 \rho a^2L^2 g^2k^4}-\frac{2s\xi v_z^2}{\pi a g}\right)^{1/2} \right] \label{eq:b0p}
\end{eqnarray}
is the critical magnetic field for the $\omega^+$ branch. When $B^+_\pm$ is complex, we set $B^+_\pm=0$.
For this branch, we have both upper and lower critical magnetic field. The former is non-negative, while the latter can be negative. The fact that $B_0$ represents the amplitude of the toroidal field requires
\begin{equation}
  B_0>0.
\end{equation}
Therefore, when $B^+_-<0$ we reset $B^+_-=0$. When $B^+_+=B^+_-=0$, this branch vanishes.
When $v_z<0$, $s=-1$. In this case $B^+_-<0$. So we reset it to be zero. 

Similarly, solving inequality~(\ref{eq:criterion}) for $B_0$, we obtain
\begin{equation}
  B^-_-<B_0<B^-_+,
\end{equation}
for the $\omega^-$ branch, where
\begin{eqnarray}
B^-_\pm &=& B_g\left[-\left(\frac{DP}{8\rho a^2L^2g^2k^4} \right)^{1/2} \right. \nonumber \\
 &&  \pm \left.\left(\frac{DP}{8 \rho a^2L^2 g^2k^4}-\frac{2s\xi v_z^2}{\pi a g}\right)^{1/2} \right]. \label{eq:b0m}
\end{eqnarray}
When $B^-_\pm$ is complex, we set $B^-_\pm=0$.  For this branch, we also have both upper and lower critical magnetic field. However, the latter is always non-positive, so we reset $B^-_-=0$. The former can be negative, too. When $B^-_+<0$ we reset $B^-_+=0$. When $B^-_+=B^-_-=0$ this branch vanishes since $B_0>0$ is required. Only when $v_z<0$ is $B^-_+$ positive.

\vspace{3mm}
\centerline{\epsfysize=8.cm \epsfbox{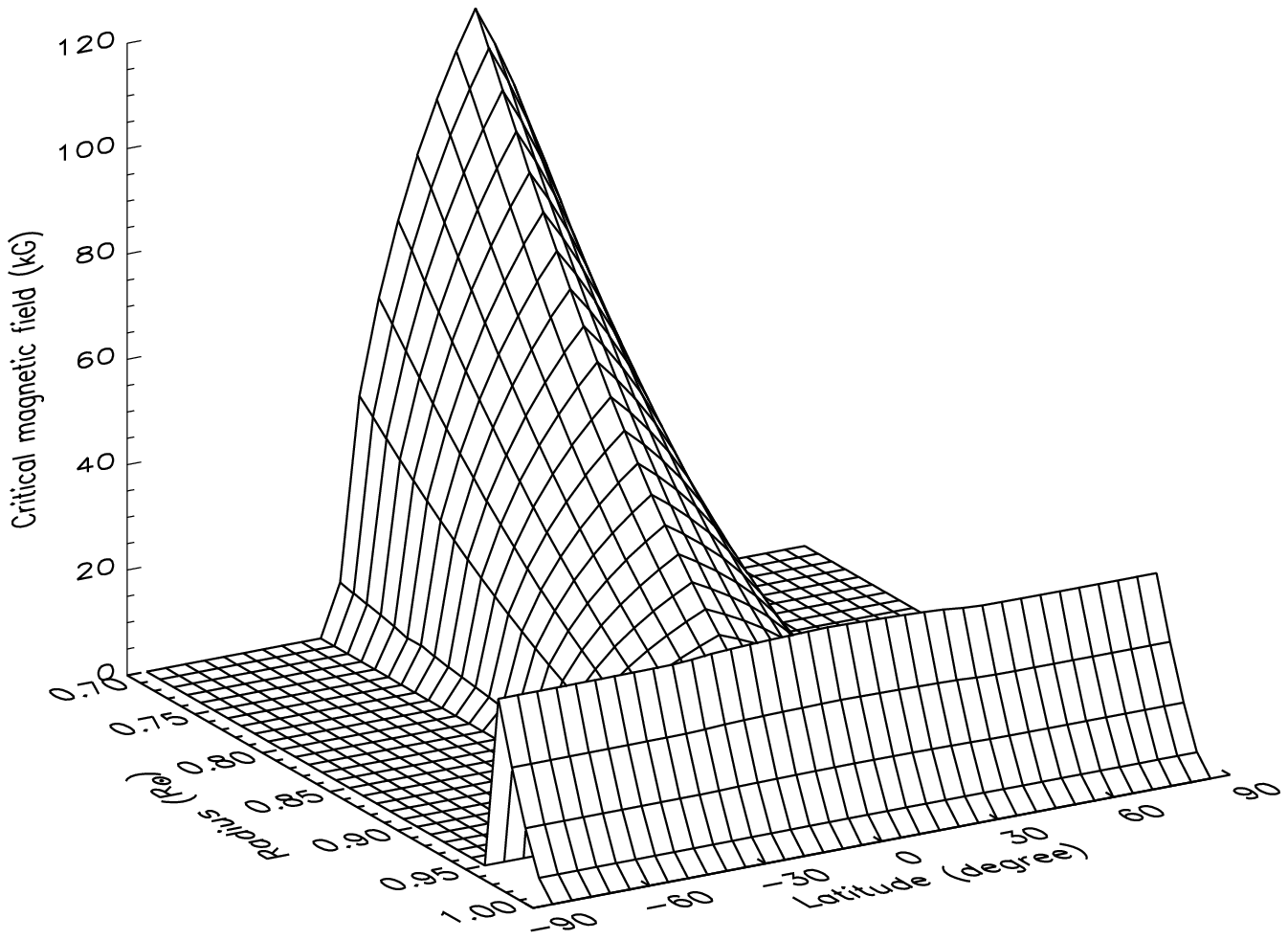}}
\figcaption[f2.eps]{
Critical magnetic field as a function of radius and latitude.
\label{fig:cri}}
\vspace{3mm}

From Eqs.~(\ref{eq:b0p}) and (\ref{eq:b0m}) we can see that the critical magnetic field of a magnetic flux tube depends on not only the velocity (including direction $s$ and speed $|v_z|$) of flows, the size (including radius $a$ and length $L$) of the flux tube, and its location, but also the fractional area $\xi$. When the $\alpha\Omega$-effect is neglected (letting $D=0$), the critical field is determined by competition between buoyant force and dynamic pressure of downflows,
\begin{equation}
  B_+^\pm = B_g(-2s\xi v_z^2/\pi a g)^{1/2}.
\end{equation}
This is equivalent to the magnetic field by equating the buoyant and dynamic acceleration given in Eq.~(\ref{eq:gb}) and Eq.~(\ref{eq:gn}). Obviously, the critical field does not vanish only when there is a downflow so that $s=-1$ and $\xi\ne 0$. In fact, sunspots are often observed near the edges of granulations where there exist downflows. The critical field is proportional to the speed of downflows, but inversely proportional to the square root of the radius of the flux tube.
Fig.~\ref{fig:cri} shows $B^\pm_+$ as a function of solar radius and latitude with $a=2000$ km, $v_z=-0.5$ km s$^{-1}$ when $r>0.95 R_\sun$, $\xi=1$, and $L=50 a$. The other parameters are the same as in Table~\ref{tbl:period}.

When the $\alpha\Omega$-effect is included, the critical magnetic field for the equatorward branch is not zero even if there is no downflow. In this case, Eq.~(\ref{eq:b0p}) reduces to
\begin{equation}
   B^+_+ = B_g\left(\frac{DP}{2\rho a^2L^2g^2k^4} \right)^{1/2}.
\end{equation}
Obviously, the thinner and shorter the flux tube the larger the critical magnetic field. The critical field without downflows is weaker than the corresponding field with downflows, but it does not vanish. The reason is that $\alpha\Omega$-effect goes against the buoyant instability. In this case, the critical field for another branch is zero.

\subsection{Growth rate}

From the dispersion relation given by Eq.~(\ref{eq:omega}), we find the growth rate for the $\omega^\pm$ branch,
\begin{eqnarray}
  \Gamma^\pm &=& -\omega^\pm_I = -(\beta_b+\beta_n)k^2 \pm \sqrt{D} \nonumber \\
&=& \frac{\sqrt{2}\,agLk^2}{c}\frac{(B^\pm_+-B_0)(B_0-B^\pm_-)}{B_g|B_0|}, \label{eq:gamma}
\end{eqnarray}
where $c=(P/\rho)^{1/2}$ is the sound speed, and $B^\pm_\pm$ are given in Eqs.~(\ref{eq:b0p}) and (\ref{eq:b0m}). Obviously, the growth rate is positive when
\begin{equation}
  B^\pm_-<B_0<B^\pm_+,
\end{equation}
zero when $B_0=B^\pm_\pm$, negative otherwise. This predicts that a magnetic flux tube experiences an ascending phase and a descending phase, implying that sunspots grow slowly, last for some time, and then disappear, since a sunspot appears when a flux tube crosses the surface of the Sun. This is in good agreement with sunspot observations.

\centerline{\epsfysize=8.cm \epsfbox{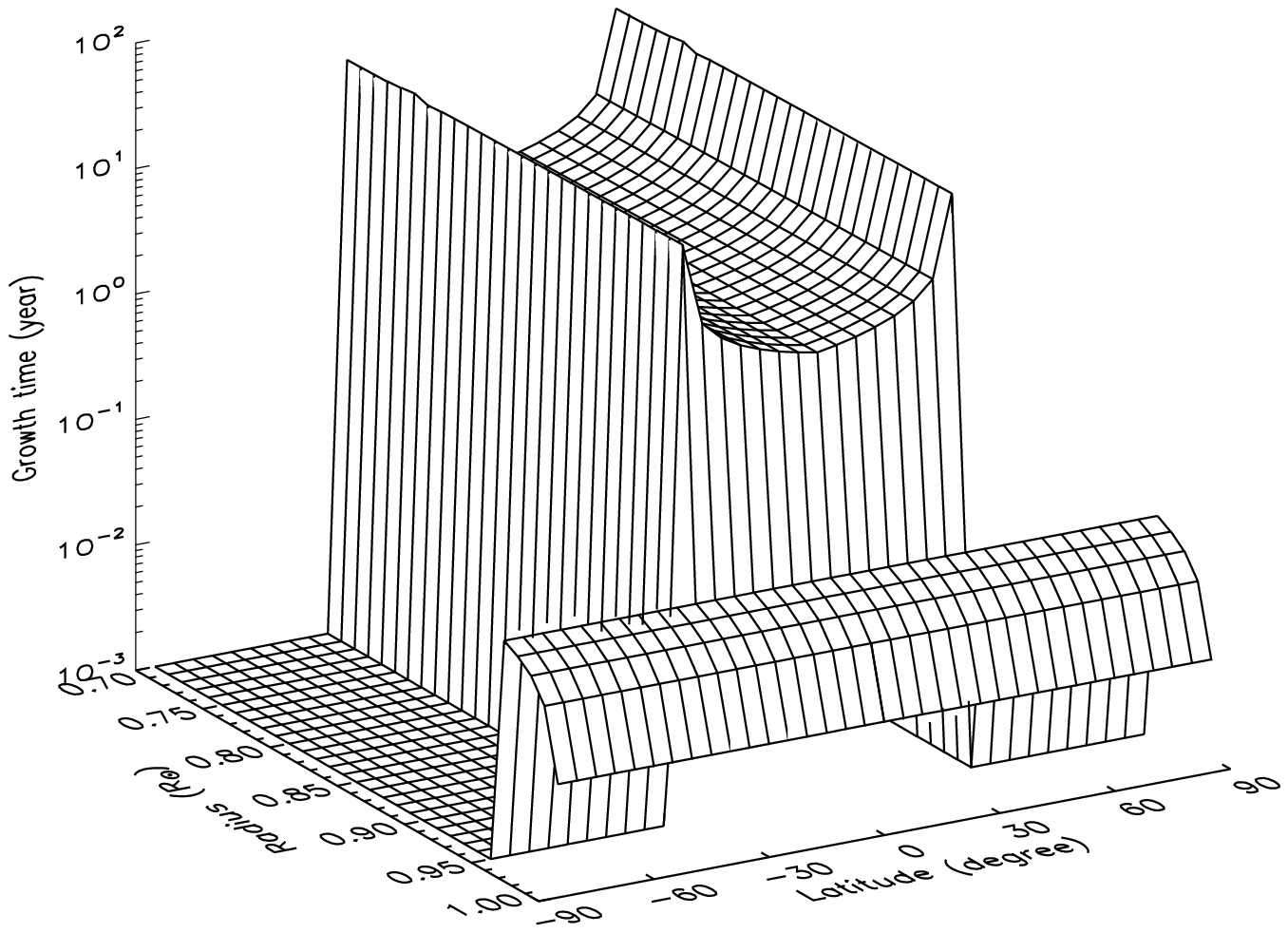}}
\figcaption[f3.eps]{
Growth time of the critical magnetic field shown in Fig.~\ref{fig:cri} as a function of radius and latitude.
\label{fig:gro}}
\vspace{3mm}

Eq.~(\ref{eq:gamma}) shows two wave branches, one that propagates equatorwards, and another that propagates polewards, can be excited. Fig.~\ref{fig:gro} shows the required time $\tau^\pm$
\begin{equation}
   \tau^\pm = \int_{B_1}^{B^\pm_+} [B\Gamma^\pm(B)]^{-1}d B \label{eq:tau}
\end{equation}
for the tube field to grow from $B_1=10^{-3} B^\pm_+$ to $B^{\pm}_+$ (given by Fig.~\ref{fig:cri}) as a function of solar radius and latitude.

When $v_z=0$, Eq.~(\ref{eq:gamma}) reduces to
\begin{equation}
  \Gamma^\pm = \pm\sqrt{D}-\beta_b k^2.
\end{equation}
This shows the magnetic field in a magnetic flux tube can grow since the $\alpha\Omega$-effect can go against the buoyant instability in the equatorward branch even if there is no downflow. Nevertheless, the growth rate is much smaller than that with downflows.

\subsection{Dynamo waves}

Solving Eq.~(\ref{eq:disp1}) for $A_0'$, we can express the temporal evolution of the toroidal magnetic field $B$ and the poloidal vector potential $A$ in terms of the initial field $B_0$ for each magnetic flux tube as follows:
\begin{eqnarray}
   B^\pm &=& B_0 e^{\Gamma^\pm t}\exp\left[i\left(\frac{2\pi t}{T}\mp\vb{k}\cdot\vb{x}\right)\right], \label{eq:bpm}\\
   A^\pm &=& B_0\frac{\alpha T}{4\pi}e^{\Gamma^\pm t}\exp\left[i\left(\frac{2\pi t}{T}\mp\vb{k}\cdot\vb{x}-  \frac{\pi}{4}\right)\right]. \label{eq:apm}
\end{eqnarray}
The wavenumber vector $\vb{k}$ should satisfy Eq.~(\ref{eq:d}). Substituting Eqs.~(\ref{eq:bpm}-\ref{eq:apm}) into Eq.~(\ref{eq:b}), we obtain,
\begin{eqnarray}
 \vb{B}^\pm &=&B_0 e^{\Gamma^\pm t}\exp\left[i\left(\frac{2\pi t}{T}\mp\vb{k}\cdot\vb{x}+\phi^\pm\right)\right] 
  \cdot\left\{\vb{e}_\phi \right. \nonumber \\
 && \left. +\frac{\alpha T}{4\pi}(-k_z\vb{e}_x+ k_x\vb{e}_z)\exp\left[-i\frac{\pi}{4}(1\pm 2)\right]\right\},
    \label{eq:wave}
\end{eqnarray}
where $k_x=\sqrt{2}\,\pi/r$, $k_z=2\pi/r$ and $k_y=0$ are required by the condition $D>0$, and $\phi^\pm$ are the initial phase angles for these two branches.

This equation shows that the toroidal field leads the poloidal field by a phase of $3\pi/4$ for the equatorward branch, but lags the poloidal field by a phase of $\pi/4$ for the poleward branch. The former is in agreement with the observations, which reveal that the poloidal component is almost in antiphase with the toroidal component. 

From Fig~\ref{fig:gro} it can be seen that the required time for the magnetic field of a typical magnetic flux tube to grow from a very low level to the critical field is very short in comparison with the cycle period when downflows are present. Using Eq.~(\ref{eq:tau}) it is easy to check that the decay time for such a magnetic flux tube is of the same order of magnitude as the growth time. Observations show that sunspots last, on average, a week or two. Therefore, our theory and observations are in agreement with each other. However, this timescale is much shorter than the cycle period. This suggests that magnetic flux tubes are only the basic elements of the magnetic flux tube wave described by Eq.~(\ref{eq:model}).
Therefore, we use the total magnetic flux $\Phi=N \pi \bar{a}^2 {\cal B}$ rather than the magnetic field strength $B$ for each magnetic flux tube as the wave  variable of the magnetic flux tube wave, where $N$ is the total number of magntic flux tubes present at the same time, $\bar{a}$ is the statistically averaged radius of magnetic flux tubes, and ${\cal B}$ is the statistically averaged magnetic field strength of magnetic flux tubes. Since $\bar{a}$ and ${\cal B}$ are mean quantities, only the number of flux tubes is variable, we finally find that the flux tube number varies with time and latitude
\begin{equation}
  N = N_0 \exp\left[i\left(\frac{2\pi t}{T}\mp\vb{k}\cdot\vb{x}+\phi^\pm\right)\right], \label{eq:psi}
\end{equation}
since $x=r[1+\tan(\pi/4-\Phi)]/\sqrt{2}$, where $\Phi$ is the latitude. $N_0$ is a constant in our single-tube theory.

For the equatorward branch, we can investigate how its peak propagates with time by setting
\begin{equation}
   2\pi t/T - k_x x -k_z z + \phi^+ = (n-1)\pi,
\end{equation}
where $n$ equals to the integer part of $1+2t/T$. Since $k_x=\sqrt{2}\pi/r$ and $k_z=2\pi/r$, 
we obtain the migration equation of the wave peak
\begin{equation}
  \Phi^+ = \frac{\pi}{4}-\tan^{-1}\left[\frac{2t}{T}-(n-1)+\tan\left(\frac{\pi}{4}-\frac{40\pi}{180}\right)\right].
   \label{eq:eb}
\end{equation}
where we have assumed $\Phi^+=40^\circ$ when $t=0$. Obviously, the integer $n$
plays a role of cycle number.

\vspace{3mm}
\centerline{\epsfysize=8.cm \epsfbox{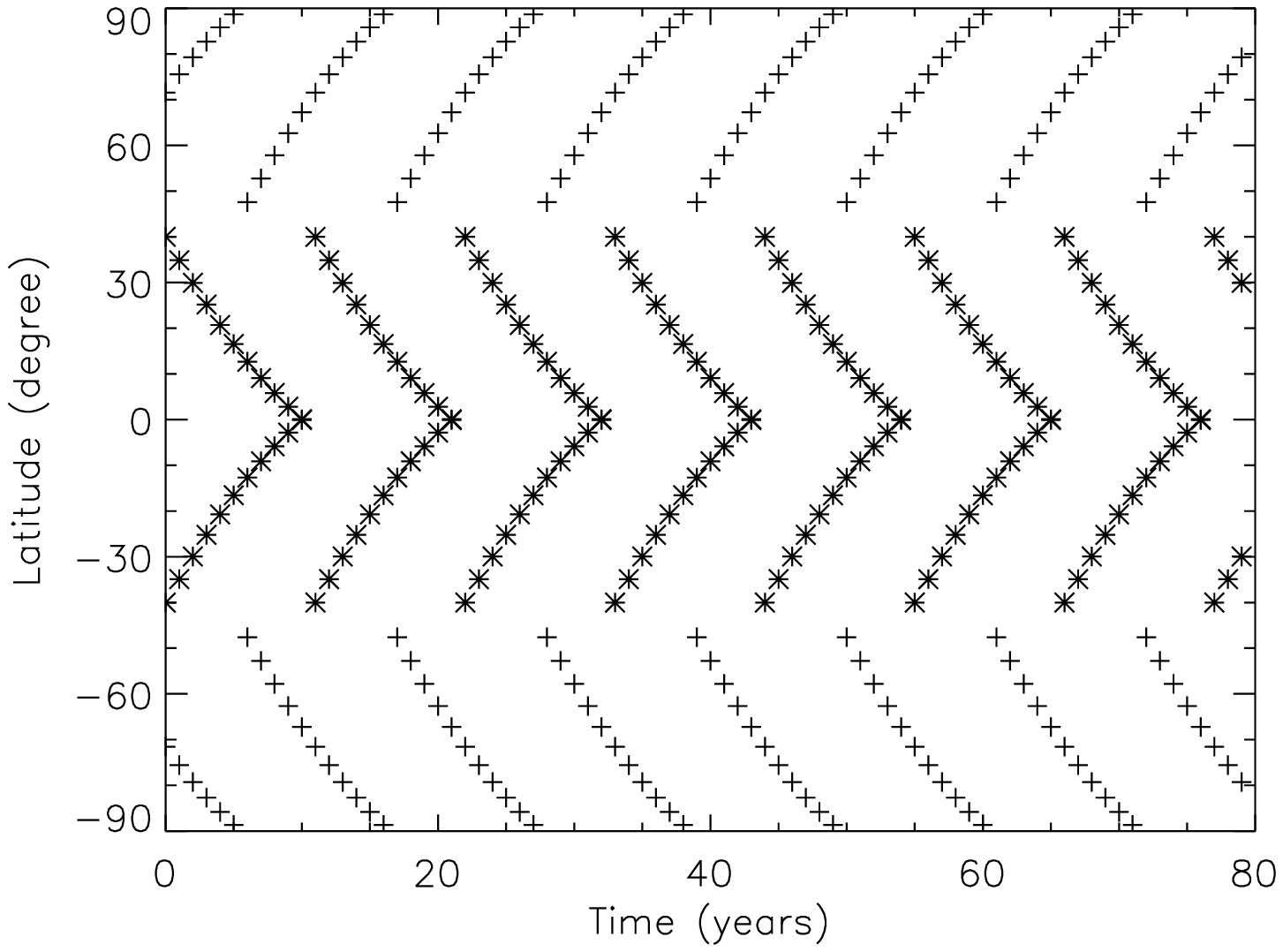}}
\figcaption[f4.eps]{
Calculated Butterfly diagram.
\label{fig:butterfly}}
\vspace{3mm}

For the poleward branch, we can only see the reflected wave because it propagates into the interior of the sun.
The reflection point is assumed to be located at that point where the gas density is one order of magnitude larger than that of the source region ($r_{\s{src}}=0.96 R_\sun$). This leads to $r_{\s{rfl}}\approx 0.86 R_\sun$. The phase velocity of the wave in the radial direction approximately equals to $100$ cm/s. Therefore, it takes about $t_0=5.5$ years for the wave to travel from the source region ($r_{\s{src}}=0.96 R_\sun$) to the reflection point ($r_{\s{rfl}}=0.86 R_\sun$), and then to the surface ($r=R_\sun$). As a result, the wave peak propagation equation for the poleward branch reads as follows:
\begin{equation}
   2\pi (t+t_0)/T + k_x x + k_z z + \phi^- = (n'-1)\pi,
\end{equation}
where $n'$ equals to the integer part of $1+2(t+t_0)/T$. The solution of this equation is
\begin{equation}
  \Phi^- = \frac{\pi}{4}+\tan^{-1}\left[\frac{2(t+t_0)}{T}-(n'-1)-\tan\left(\frac{\pi}{4}-\frac{45\pi}{180}\right)\right],
   \label{eq:pb}
\end{equation}
where we have assumed $\Phi^-=45^\circ$ when $t+t_0=0$. This implies a phase lag of $\pi/2$ with respect to the equatorward branch, as shown in Fig.~\ref{fig:butterfly}, which plots Eqs.~(\ref{eq:eb}) and (\ref{eq:pb}).

Since $N_0$ is a constant in our single-tube theory, we predict that the total number of flux tubes, on average, does not vary with time. This is in conflict with observations, which show that the sunspot number increases in the ascending phase, and then decrease in the descending phase of the solar cycle. This problem may be solved by taking into account interactions between magnetic flux tubes.

\section{Comparison with observations}\label{sect:s3}

In this section we try to interpret the relevant observational facts $1- 10$ listed in \S\ref{sect:s0} in order, using the results obtained in the previous section.

\begin{enumerate}
\item From Table~\ref{tbl:period} we can see that the cycle period equals to the observed 22 years when the dynamo operates above $r=0.95 R_\sun$, as inferred from helioseismology by Antia et al. (2000). Eqs.~(\ref{eq:period}) and (\ref{eq:d}) show that the cycle period is determined by the internal differential rotation and the cyclonic turbulent convection motions in the convection zone.
\item Fig.~\ref{fig:butterfly} shows that the calculated butterfly diagram is in good agreement with the observed extended butterfly diagram by Makarov \&\ Sivaraman (1989). 
\item What we observe at high latitudes are the reflected waves. They must be weaker than the original waves since the relfection is not necessarily complete. This naturally explains why the poleward branch is weaker than the equatorward branch. In other words, strong activity should be confined to low heliographic latitudes $|\Phi|\le 35^\circ$ of the equatorward branch, as observed.
\item Phase dilemmas:
\begin{enumerate}
  \item Eq.~(\ref{eq:wave}) shows that the toroidal field leads the poloidal field by a phase of $3\pi/4$ for the equatorward branch. This is in good agreement with the observation: the poloidal field is almost in antiphase with the toroidal component;
 \item The $\pi/2$ phase lag of the poleward branch with respect to the equatorward branch can be explained by the time lag 
of 5.5 years for the former to travel from the source region to the reflection point, and then from the reflection point to the surface.
 \end{enumerate}
\item From Eqs.~(\ref{eq:b0p}), (\ref{eq:b0m}), and Fig.~\ref{fig:cri}, we can see that the allowed magnetic field (i.e., the critical field) in a concentrated flux tube (specified by cross section radius $a$ and tube length $L$) is strong at the depth of $r=0.96 R_\sun$. Fig~\ref{fig:gro} shows that the growth (or decay) time for a typical flux tube is about several weeks at the depth.
\item We use the internal rotation rates obtained from helioseismology (Fig.~\ref{fig:rot}).
\item Fig.~\ref{fig:cri} shows that the magnetic fields peak at $r=0.96 R_\sun$ when high-speed downflows are assumed to occur above $r=0.95 R_\sun$. Though the critical field is stronger at the base of the convection zone, the growth time is  longer than the cycle period.
\end{enumerate}

\vspace{3mm}
\centerline{\epsfysize=10.cm \epsfbox{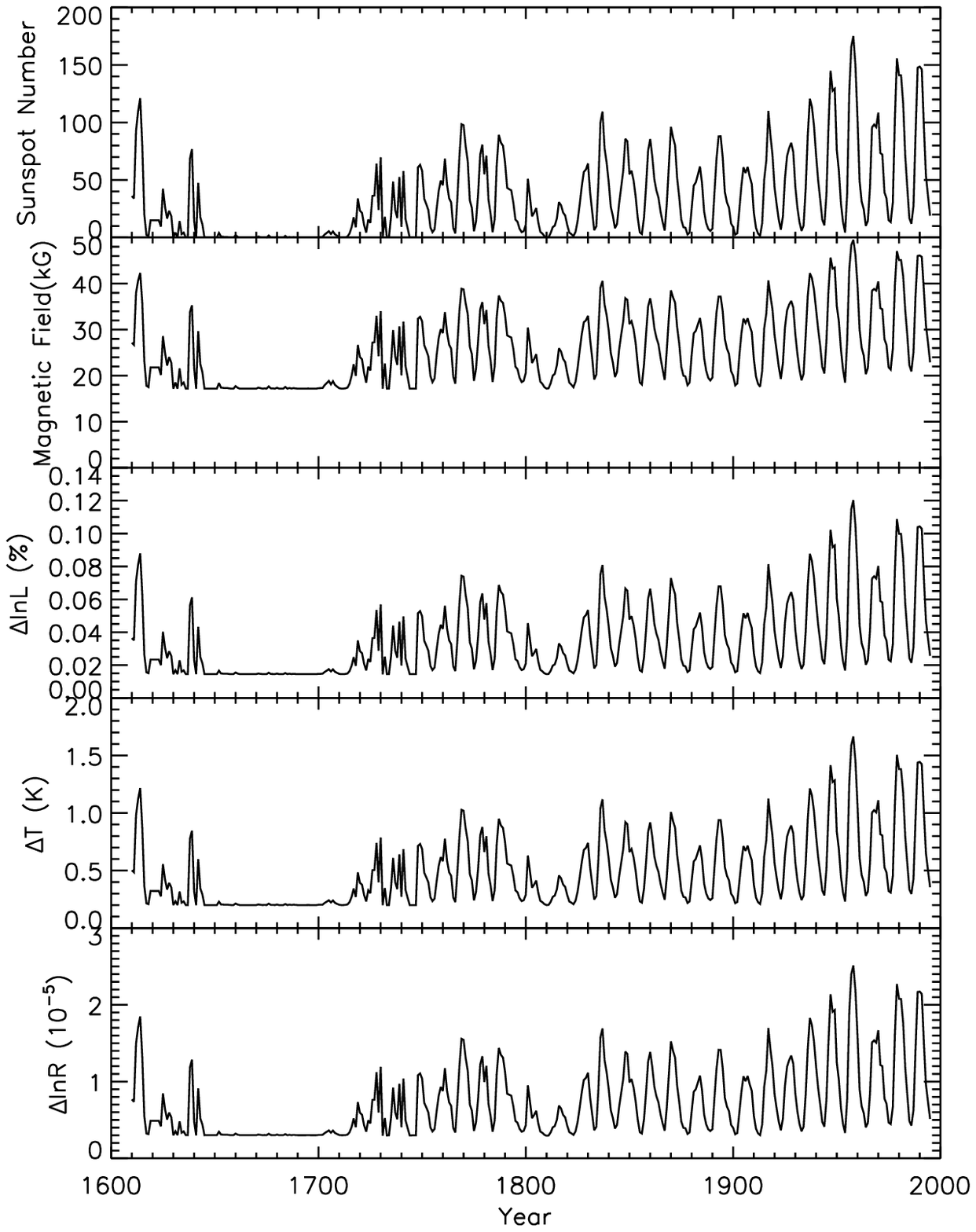}}
\figcaption[f5.eps]{
Cyclic variations of solar global parameters.
\label{fig:cyc}}
\vspace{3mm}

\centerline{\epsfysize=6.cm \epsfbox{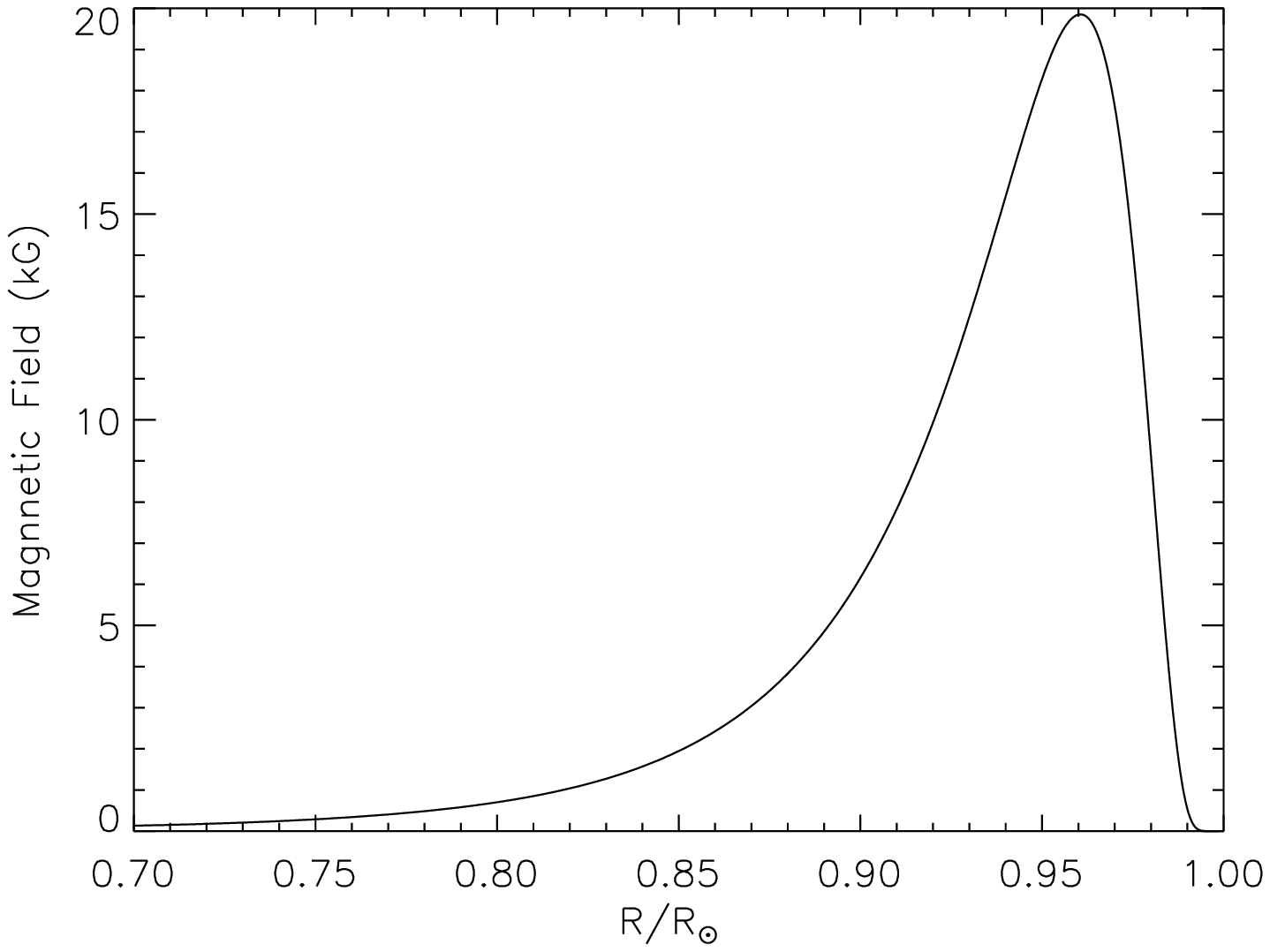}}
\figcaption[f6.eps]{
Assumed solar internal magnetic field as a function of radius in 1996, as suggested by helioseismology \cite{ACT00}.
\label{fig:mag}}
\vspace{3mm}

As elucidated by Li \& Sofia (2001) and Sofia \&\ Li (2001), the cyclic variations of global solar parameters (Fig.~\ref{fig:cyc} ) can be reproduced by the magnetic field peaked at the depth of $r=0.96 R_\sun$ depicted in Fig.~\ref{fig:mag}:
\begin{enumerate}
\setcounter{enumi}{7}
\item The first and third panels of Fig.~\ref{fig:cyc} show that the total solar luminosity varies in phase with the solar cycle. The amplitude equals about $0.1\%$.
\item The first and fourth panels of Fig.~\ref{fig:cyc} show that the solar effective temperature varies in phase with the solar cycle. The amplitude equals about $1.5^\circ C$.
\item The first and fifth panels of Fig.~\ref{fig:cyc} show that the solar radius varies in phase with the solar cycle. The  amplitude equals about 20 mas.
\end{enumerate}
If the magnetic field is placed at the base of the convection zone, we cannot find a strongth that simultaneously reproduce the observed cyclic variations for these three global solar parameters.

\section{Conclusion}\label{sect:s4}

We use 10 main observational facts pertinent to the solar cycle activity to filter the CZ, OL, IF and BL solar cycle models. We find that none of them are satisfactory. In particular, recent helioseimic inversions \cite{ACT00} that determine the location of solar internal magnetic fields at $r=0.96 R_\sun$, rule out the OL, IF, and BL models, which place the solar magnetic fields at the base of the convection zone. This is one of the main motivations to reconsider the CZ dynamos. The second motivation comes from trying to reproduce the observed cyclic variations of total solar irradiance \cite{FL98}, solar effective temperature \cite{GL97}, and solar radius \cite{EKBS00}. Extensive numerical experiments \cite{LS01} shows that only magnetic fields located in the convection zone can simultaneously produce the observed cyclic variations of these three global solar parameters. The third motivation originates from numerical simulations of the solar convection zone \cite{CS89,N99} because they indicate a major presence of downward-moving plumes of high velocity.

For the CZ dynamos, we have to face the magnetic buoyancy instability problem. The best way to do so is to incorporate the buoyancy into the dynamo euqations. This forces us to treat the mean field in a flux tube. To include the dynamic pressure of a flow, the flux tube should be considered to be a body. As a field, both the cyclonic convection ($\alpha$-effect) and differential rotation ($\Omega$-effect) play a role. As a body, the tube experiences not only a buoyant force, but also a dynamic pressure of downflows above the tube. We show that these two dynamic effects can be incorporated into the dynamo equations by adding to them two diffusion terms.

We analyze and solve the extended dynamo equations in the linear approximation by using the observed solar internal rotation rates and assuming a downflow suggested by numerical simulations of the solar convection zone. The results are in agreement with all the properties of the solar cycle listed in \S\ref{sect:s0}.

We can not only interpret the observational facts of the 22-year cycle, but also make interesting predictions. For example, Figs.~\ref{fig:cri} and \ref{fig:gro} may imply longer activity periods of dacades to centuries in addition to the 22-year cycle, as suggested by the historical data. We'll explore them separately.

\acknowledgements
We want to thank Dr. Sarbani Basu for her kindly providing us with rotation rate data, Dr. Frank J. Robinson for his help, and Prof. Pierre Demarque for useful discussions. This work was supported in part by a grant from the National Aeronautics and Space Administration, and in part by Natural Science Foundation of China (project 19675064).

\begin{deluxetable}{lllllllll|rrrrrrrrrr}
\tabletypesize{\scriptsize}
\tablecaption{The length of the solar cycle (unit: year) as a function of radius and latitude. \label{tbl:period}}
\tablewidth{0pt}
\tablehead{
   \multicolumn{1}{l}{R} & \multicolumn{8}{c}{$k_x=\frac{\sqrt{2}\,\pi}{r}$,\ $k_z=\frac{2\pi}{r}$,\ $\alpha=112$ cm s$^{-1}$} & \multicolumn{10}{c}{$k_x=-\frac{\sqrt{2}\,\pi}{r}$,\ $k_z=-\frac{2\pi}{r}$,\ $\alpha=6.4$ cm s$^{-1}$}  }  
\startdata
     &   0$^\circ$ &   5$^\circ$ &  10$^\circ$ &  15$^\circ$ &  20$^\circ$ &  25$^\circ$ &  30$^\circ$ &  35$^\circ$ 
  &  40$^\circ$ &  45$^\circ$ &  50$^\circ$ &  55$^\circ$ &  60$^\circ$ &  65$^\circ$ &  70$^\circ$ &  75$^\circ$ &  80$^\circ$ &  85$^\circ$ \\ \tableline 
0.72 &   1 &   1 &   1 &   2 &   2 &   3 &   4 &  19 &  19 &  13 &  11 &   9 &   8 &   8 &   7 &   7 &   7 &   6 \\
0.73 &   1 &   1 &   1 &   2 &   2 &   3 &   4 &  19 &  19 &  13 &  11 &  10 &   9 &   8 &   8 &   7 &   7 &   7 \\
0.74 &   1 &   1 &   1 &   2 &   2 &   3 &   4 &  20 &  19 &  13 &  11 &  10 &   9 &   8 &   8 &   7 &   7 &   7 \\
0.75 &   1 &   1 &   1 &   2 &   2 &   3 &   4 &  21 &  19 &  13 &  11 &  10 &   9 &   8 &   8 &   7 &   7 &   7 \\
0.76 &   1 &   1 &   1 &   2 &   2 &   3 &   4 &  19 &  19 &  14 &  11 &  10 &   9 &   8 &   8 &   7 &   7 &   7 \\
0.77 &   1 &   1 &   2 &   2 &   2 &   3 &   4 &  19 &  19 &  14 &  11 &  10 &   9 &   8 &   8 &   7 &   7 &   7 \\
0.78 &   1 &   1 &   2 &   2 &   2 &   3 &   4 &  20 &  20 &  14 &  11 &  10 &   9 &   8 &   8 &   7 &   7 &   7 \\
0.79 &   1 &   1 &   2 &   2 &   2 &   3 &   4 &  20 &  20 &  14 &  11 &  10 &   9 &   8 &   8 &   8 &   7 &   7 \\
0.80 &   1 &   1 &   2 &   2 &   2 &   3 &   4 &  20 &  20 &  14 &  11 &  10 &   9 &   8 &   8 &   8 &   7 &   7 \\
0.81 &   1 &   1 &   2 &   2 &   2 &   3 &   4 &  20 &  20 &  14 &  11 &  10 &   9 &   8 &   8 &   8 &   7 &   7 \\
0.82 &   1 &   1 &   2 &   2 &   2 &   3 &   4 &  20 &  20 &  14 &  12 &  10 &   9 &   8 &   8 &   8 &   7 &   7 \\
0.83 &   1 &   1 &   2 &   2 &   2 &   3 &   4 &  20 &  20 &  14 &  12 &  10 &   9 &   9 &   8 &   8 &   7 &   7 \\
0.84 &   1 &   1 &   2 &   2 &   2 &   3 &   4 &  20 &  20 &  14 &  12 &  10 &   9 &   9 &   8 &   8 &   8 &   7 \\
0.85 &   1 &   1 &   2 &   2 &   2 &   3 &   4 &  20 &  20 &  14 &  12 &  10 &   9 &   9 &   8 &   8 &   8 &   7 \\
0.86 &   1 &   1 &   2 &   2 &   2 &   3 &   4 &  20 &  20 &  14 &  12 &  10 &   9 &   9 &   8 &   8 &   8 &   7 \\
0.87 &   1 &   1 &   2 &   2 &   2 &   3 &   4 &  21 &  21 &  14 &  12 &  10 &   9 &   9 &   8 &   8 &   8 &   8 \\
0.88 &   1 &   1 &   2 &   2 &   2 &   3 &   4 &  21 &  21 &  14 &  12 &  10 &   9 &   9 &   8 &   8 &   8 &   8 \\
0.89 &   1 &   1 &   2 &   2 &   2 &   3 &   4 &  21 &  21 &  15 &  12 &  10 &  10 &   9 &   8 &   8 &   8 &   8 \\
0.90 &   1 &   1 &   2 &   2 &   2 &   3 &   4 &  21 &  21 &  15 &  12 &  11 &  10 &   9 &   8 &   8 &   8 &   8 \\
0.91 &   1 &   1 &   2 &   2 &   2 &   3 &   4 &  21 &  21 &  15 &  12 &  11 &  10 &   9 &   9 &   8 &   8 &   8 \\
0.92 &   1 &   1 &   2 &   2 &   2 &   3 &   4 &  21 &  21 &  15 &  12 &  11 &  10 &   9 &   9 &   8 &   8 &   8 \\
0.93 &   1 &   2 &   2 &   2 &   2 &   3 &   4 &  21 &  21 &  15 &  12 &  11 &  10 &   9 &   9 &   8 &   8 &   8 \\
0.94 &   1 &   2 &   2 &   2 &   2 &   3 &   4 &  21 &  21 &  15 &  12 &  11 &  10 &   9 &   9 &   8 &   8 &   8 \\
0.95 &   1 &   2 &   2 &   2 &   2 &   3 &   4 &  21 &  21 &  15 &  12 &  11 &  10 &   9 &   9 &   8 &   8 &   8 \\
0.96 &   1 &   2 &   2 &   2 &   2 &   3 &   4 &  22 &  22 &  15 &  12 &  11 &  10 &   9 &   9 &   8 &   8 &   8 \\
0.97 &   1 &   2 &   2 &   2 &   2 &   3 &   4 &  22 &  22 &  15 &  12 &  11 &  10 &   9 &   9 &   8 &   8 &   8 \\
0.98 &   1 &   2 &   2 &   2 &   2 &   3 &   4 &  22 &  22 &  15 &  13 &  11 &  10 &   9 &   9 &   9 &   8 &   8 \\
0.99 &   1 &   2 &   2 &   2 &   2 &   3 &   5 &  22 &  22 &  15 &  13 &  11 &  10 &  10 &   9 &   9 &   9 &   9 \\
1.00 &   1 &   2 &   2 &   2 &   2 &   3 &   5 &  22 &  22 &  16 &  13 &  11 &  10 &  10 &   9 &   9 &   9 &   9 \\
\enddata
\end{deluxetable}

\end{document}